\pdfoutput=1
\documentclass[PAPER, atlasdraft=false, UKenglish, cernpreprint, texmf, orcidlogo]{atlasdoc}
\usepackage{atlaspackage}
\usepackage{atlasbiblatex}

\usepackage{atlasphysics}
\usepackage{multirow}
\graphicspath{{logos/}{figures/}}

\AtlasTitle{Search for Higgs boson decays into a $Z$ boson and a light hadronically decaying resonance in $pp$ collisions at $\sqrt{s}$=13 TeV with the ATLAS detector}

\AtlasAbstract{%
A search for decays of the Higgs boson into a $Z$ boson and a light resonance, with a mass of 0.5--3.5~GeV, is performed using the full 140~fb$^{-1}$ dataset of 13~TeV proton--proton collisions recorded by the ATLAS detector during LHC Run~2. Leptonic decays of the $Z$ boson and hadronic decays of the light resonance are considered. The resonance can be interpreted as a $J/\psi$ or $\eta_c$ meson, an axion-like particle, or a light pseudoscalar predicted in two-Higgs-doublet models. Due to its low mass, this resonance is produced with a high Lorentz boost in the laboratory frame and therefore reconstructed as a single small-radius jet of hadrons. A neural network is used to correct the Monte Carlo simulation of the total expected background using data from sideband regions. Two additional neural networks are used to distinguish signal from background, enhancing the purity of the signal region. A binned profile-likelihood fit is performed on the final-state invariant mass distribution. No significant excess of events relative to the expected background is observed, and upper limits at 95\% confidence level are set on the Higgs boson's branching fraction to a $Z$ boson and a light resonance. The exclusion limit is $\sim$10\% for the lower masses, and increases for higher masses. Upper limits on the effective coupling $C^\text{eff}_{ZH}/\Lambda$ of an axion-like particle to a Higgs boson and $Z$ boson are also set at 95\% confidence level, and range from 0.9 to 2~TeV$^{-1}$.}

\AtlasRefCode{HDBS-2021-09}

\PreprintIdNumber{CERN-EP-2024-261}

\HepDataRecord{https://www.hepdata.net/record/153859}

\AtlasJournalRef{Phys.\ Lett.\ B. 868 (2025) 139671}
\AtlasDOI{10.1016/j.physletb.2025.139671}

\AtlasRefCode{HDBS-2021-09}
\hypersetup{pdftitle={ATLAS document},pdfauthor={The ATLAS Collaboration}}

\begin{document}

\maketitle

\tableofcontents

\section{Introduction}
\label{sec:intro}

Since its discovery in 2012~\cite{HIGG-2012-27,CMS-HIG-12-028}, the Higgs boson has been studied intensively by the ATLAS and CMS experiments at the CERN Large Hadron Collider (LHC)~\cite{Evans:2008zzb}. So far, all of its measured properties have been found to be consistent with the  predictions of the Standard Model (SM)~\cite{HIGG-2021-23,CMS-HIG-22-001} and no decays to unknown particles have been observed. Currently the most restrictive upper limits at 95\% confidence level (CL) on the Higgs branching fractions to undetected (invisible) decays are 16\% (16\%)~\cite{CMS-HIG-22-001} from CMS and 12\% (11\%)~\cite{HIGG-2021-23, HIGG-2021-05} from ATLAS. However, chasing beyond the SM (BSM) Higgs decays is compelling because several models predict additional particles that might couple with the Higgs boson. These particles address limitations of the SM such as the nature of dark matter~\cite{Arcadi_2020_} or the strong CP problem~\cite{PhysRevLett.38.1440}. Because the Higgs boson decay width is predicted to be so small, (4.1 MeV~\cite{deFlorian:2016spz}), even a small coupling to a non-SM particle could result in new decay modes with large enough branching fractions to be experimentally observed at the LHC.

One of the simplest extensions of the SM is the two-Higgs-doublet model (2HDM) ~\cite{Branco:2011iw,Gunion:2002zf}, which introduces a second Higgs doublet, allowing the presence of additional lighter Higgs bosons.  For some parameter values in both the 2HDM and the 2HDM with an additional singlet (2HDM$+$S)~\cite{Muhlleitner:2016mzt,Chen:2013jvg}, the additional particles can have large couplings to the observed Higgs boson. Specific parameter values of the 2HDM produce the Minimal Supersymmetric Standard Model (MSSM)~\cite{Csaki:1996ks}, as well as its extension, called next-to-MSSM~\cite{MANIATIS_2010}, with an additional scalar field, which can solve the $\mu$-problem of the MSSM~\cite{Babu:2001gp}, and greatly reduce the fine-tuning and little-hierarchy problems.

Axions and axion-like particles (ALPs) are other candidates for light resonances that can couple to the Higgs boson~\cite{Baker:2013zta, Bauer:2017ris}. While axions were initially introduced to address the strong CP problem~\cite{PhysRevLett.38.1440}, ALPs are currently of
much wider interest. They are one of the favourite dark-matter candidates~\cite{Chatrchyan:2023cmz} in several cosmological models and could also explain the observed tension between the theoretical prediction and the measured magnetic moment of the muon~\cite{Buen-Abad:2021fwq}.

There has recently been increasing interest in light-resonance ($a$) decays of the Higgs boson. Both ATLAS and CMS have published several results, most of them focusing on Higgs decays to pairs of $a$ resonances ($H\to aa$), with masses of $\mathcal{O}$(1-10~\GeV) and leptonic or photonic decays of the $a$-resonance~\cite{CMS-HIG-22-003, HDBS-2019-19, EXOT-2016-22, HDBS-2021-07, HDBS-2019-09, CMS-HIG-21-003, CMS-HIG-22-007}.

However, in both the 2HDM(+S) and axion models, depending on the model parameters, the $a$ particle can have a significant hadronic branching fraction. For instance, in the type-II 2HDM+S, assuming $\tan\beta=1$, the branching fraction $\mathcal{B}(a\to gg)$ exceeds 85\% for masses up to 3~\GeV~\cite{Curtin:2013fra}. Similarly, for an ALP with coupling constants (Wilson coefficients) equal to 1, the branching fraction for decays into gluons is almost 100\% for masses greater than 3~\GeV, and  decays into pions are significant for masses up to 1~\GeV~\cite{Bauer:2017ris}.

Light scalar resonances, with mass of up to a few \GeV, originating from a Higgs boson decay, are also exist within the SM. For example, $c\bar{c}$ mesons such as $\eta_c$ and $J/\psi$ (Q) have masses ${\sim}3$~\GeV and decay mostly into hadrons, with $\mathcal{B} \sim 85\%$. However, their production through $H\to ZQ$ is suppressed and the expected branching fractions are of the order of $10^{-5}$ and $10^{-6}$ for $\eta_c$ and $J/\psi$, respectively~\cite{Isidori:2013cla}.

This letter focuses on the $H\rightarrow Za$ decay, utilizing leptonic decays of the $Z$ boson for event triggering~\cite{HDBS-2019-09}. It specifically targets a low-mass (${\leq}4$~\GeV) hadronically decaying $a$-resonance from the $H\rightarrow Za$ process. It supersedes the initial version of the analysis published in Ref.~\cite{HDBS-2018-37} and aims to address the primary limitation of that study:\ the background modelling uncertainty. The present analysis incorporates a novel neural-network approach to improve the background modelling, leading to significantly stronger exclusion limits, and includes several additional methodological upgrades. Moreover, an axion interpretation is included, resulting in the exclusion of a range of Higgs-$Z$-$a$ effective coupling values.

\section{ATLAS detector}
\label{sec:detector}
\newcommand{\AtlasCoordFootnote}{%
ATLAS uses a right-handed coordinate system with its origin at the nominal interaction point (IP)
in the centre of the detector and the \(z\)-axis along the beam pipe.
The \(x\)-axis points from the IP to the centre of the LHC ring,
and the \(y\)-axis points upwards.
Polar coordinates \((r,\phi)\) are used in the transverse plane,
\(\phi\) being the azimuthal angle around the \(z\)-axis.
The pseudorapidity is defined in terms of the polar angle \(\theta\) as \(\eta = -\ln \tan(\theta/2)\) and is equal to the rapidity
$ y = \frac{1}{2} \ln \left( \frac{E + p_z}{E - p_z} \right) $ in the relativistic limit.
Angular distance is measured in units of \(\Delta R \equiv \sqrt{(\Delta y)^{2} + (\Delta\phi)^{2}}\).}

The ATLAS experiment~\cite{PERF-2007-01} at the LHC is a multipurpose particle detector
with a forward--backward symmetric cylindrical geometry and a near \(4\pi\) coverage in
solid angle.\footnote{\AtlasCoordFootnote}
It consists of an inner tracking detector surrounded by a thin superconducting solenoid
providing a \qty{2}{\tesla} axial magnetic field, electromagnetic and hadronic calorimeters, and a muon spectrometer.
The inner tracking detector covers the pseudorapidity range \(|\eta| < 2.5\).
It consists of silicon pixel, silicon microstrip, and transition radiation tracking detectors.
Lead/liquid-argon (LAr) sampling calorimeters provide electromagnetic (EM) energy measurements
with high granularity within the region \(|\eta|< 3.2\).
A steel/scintillator-tile hadronic calorimeter covers the central pseudorapidity range (\(|\eta| < 1.7\)).
The endcap and forward regions are instrumented with LAr calorimeters
for EM and hadronic energy measurements up to \(|\eta| = 4.9\).
The muon spectrometer surrounds the calorimeters and is based on
three large superconducting air-core toroidal magnets with eight coils each.
The field integral of the toroids ranges between \num{2.0} and \qty{6.0}{\tesla\metre}
across most of the detector.
The muon spectrometer includes a system of precision tracking chambers up to \(|\eta| = 2.7\) and fast detectors for triggering up to \(|\eta| = 2.4\).
The luminosity is measured mainly by the LUCID--2~\cite{LUCID2} detector, which is located close to the beampipe.
A two-level trigger system is used to select events~\cite{TRIG-2016-01}.
The first-level trigger is implemented in hardware and uses a subset of the detector information
to accept events at a rate below \qty{100}{\kHz}.
This is followed by a software-based trigger that
reduces the accepted event rate to \qty{1}{\kHz} on average
depending on the data-taking conditions.
A software suite~\cite{SOFT-2022-02} is used in data simulation, in the reconstruction
and analysis of real and simulated data, in detector operations, and in the trigger and data acquisition
systems of the experiment.

\section{Data and simulated event samples}

\label{sec:DataAndMC}

This search uses the entire \RunTwo proton--proton $(pp)$ collision dataset recorded by the ATLAS detector at the LHC between 2015 and 2018.
The dataset corresponds to an integrated luminosity of $140~\text{fb}^{-1}$~\cite{DAPR-2021-01} at a $pp$ centre-of-mass energy of $\sqrt{s}=13~\TeV$. The data are required to satisfy criteria ensuring that the detector was in good operating condition \cite{DAPR-2018-01}.

Samples of Monte Carlo (MC) simulated signal events ($gg \to H\to Za \to \ell\ell{gg}~\text{or}~\ell\ell{q}\bar{q}$) are generated via the gluon--gluon fusion (ggF) process using \POWHEGBOX[v2]~\cite{Nason:2004rx,Frixione:2007vw,Alioli:2010xd} and the \PDFforLHC[15nnlo] next-to-next-to-leading-order (NNLO) parton distribution function (PDF) set~\cite{Butterworth:2015oua}. The simulation achieves NNLO accuracy for arbitrary inclusive $gg \rightarrow H $ observables by reweighting the Higgs boson rapidity spectrum in H$_J-$MiNLO~\cite{Hamilton:2012rf, Hamilton:2012np, Campbell:2012am} to that in HNNLO~\cite{Catani:2007vq}. The samples are normalized to match the total Higgs boson production cross-section of 55.6~pb (for a mass of 125.09~\GeV) at $\sqrt{s}=13$~\TeV~\cite{Catani:2007vq} because the event acceptance for ggF production is similar  to that for other Higgs boson production processes. %

Particle decays, hadronization, parton showers and the underlying event in the signal samples are modelled with \PYTHIA[8.212]~\cite{Sjostrand:2014zea} and \EVTGEN[1.6.0]~\cite{Lange:2001uf}, using a set of tuned parameters called the A14 tune~\cite{ATL-PHYS-PUB-2014-021} and the \NNPDF[23lo] PDF set~\cite{Ball:2012cx}. The branching fractions of the $a$ resonance are determined using the \PYTHIA[8] MSSM scenario with two Higgs doublets~\cite{Csaki:1996ks}, which predicts $a\to gg$ to be the dominant decay mode until $a\to c\bar c$ becomes kinematically accessible.
The signal MC samples used in this analysis have $a$-resonance masses covering the range 0.5--4~\GeV. The $Z$~boson is required to decay into a pair of electrons, muons, or $\tau$-leptons. Two dedicated samples with $\eta_c$ or $J/\psi$ decays ($H\to Z\;\!\eta_c~\text{or}~Z\;\!J/\psi$) are also simulated using \PYTHIA[8] and \EVTGEN. An alternative set of signal samples, where \HERWIG[7.1] \cite{Bellm:2017bvx} is used to model the parton shower and hadronization, is produced to estimate the uncertainties associated with the modelling of the $a\to$hadrons decay. %

The background in this analysis is dominated by $Z(\ell\ell)$\,+\,jets events. The Drell–Yan production is modelled in the same way as the signal samples, using \POWHEG (NLO QCD) for the matrix element generation and $\PYTHIA{+}\EVTGEN$ for parton shower and hadronization. The \CTEQ[6L1] PDF set~\cite{Pumplin:2002vw} and the AZNLO tune of \PYTHIA[8] are used. The $Z$\,+\,jets production is also modelled with \SHERPA[2.2.1]~\cite{Bothmann:2019yzt}, using the \NNPDF[3.0nnlo] PDF set~\cite{Ball:2014uwa} for the modelling of the hard interaction and parton shower. In this case, the inclusive production cross-sections are known to NNLO in QCD~\cite{Anastasiou:2003ds}. The \POWHEG sample is chosen as the nominal, because of its larger statistical power than the \SHERPA one. Only events with $Z\rightarrow ee$ or $Z\rightarrow \mu\mu$ decays are considered. An additional $\sim$5\% contribution is expected from $tau$ leptonic decays from $Z\rightarrow \tau\tau$. However, this contribution is suppressed by the $Z$-mass requirement described in Sec.~\ref{sec:ES} and the remaining events are included in the background prediction described in Sec.~\ref{sec:RW}.

The $ZZ$ and $ZW$ processes constitute small (${<}1\%$) backgrounds in this analysis. They are modelled %
with \SHERPA[2.2.1], using the \NNPDF[3.0nnlo] PDF set. The $t\bar t$ process also produces a small (${<}1\%$) background in this analysis.  For the $t\bar t$ background, the hard interaction is modelled using \POWHEG, while the decay, hadronization, parton shower and underlying event are modelled using \PYTHIA[8] and \EVTGEN. The above-mentioned processes constitute the total background in this analysis.

The simulated event samples also include a $\GEANT$-based simulation of the ATLAS detector and its response~\cite{Agostinelli:2002hh,SOFT-2010-01}, and the effect of other $pp$ interactions in either the bunch crossing containing the hard interaction or neighbouring ones.

\section{Event selection}
\label{sec:ES}

The low-mass signal resonance is highly boosted due to the relatively large kinetic energy imparted to it because of the sizeable mass difference between the Higgs boson and $Z$ boson. Its hadronic decay products are therefore reconstructed as a single small-radius ($R = 0.4$) jet  using particle flow objects \cite{PERF-2015-09} as input to the \antikt algorithm~\cite{Cacciari:2008gp,Fastjet}.

Jets are calibrated such that the average detector-level jet energy scale (JES) matches that of the corresponding particle-level jets, using a combination of simulation-based and in situ techniques \cite{JETM-2018-05}.  They are required to have $\pT>20~\GeV$ and $\vert\eta\vert<2.5$, and satisfy a jet cleaning requirement~\cite{ATLAS-CONF-2015-029} to suppress non hard collision jets . To reject jets from \pileup interactions, jets with $\pT<60~\GeV$ and $\vert\eta\vert<2.4$ are required to pass a \textit{Tight} \enquote{jet vertex tagger}~\cite{PERF-2014-03} requirement.

The signal exhibits a two-prong structure being a light resonance decaying to two-gluons or two-quarks, while the background is typically a single prong or gluon splitting (jet associated to the production of a Z-boson). Therefore, jet substructure variables are used to discriminate between them. The variables are built using tracks matched to the calorimeter jet by ghost association~\cite{Cacciari:2007fd}. In this method the tracks are included in the jet clustering process with negligible energy so that they do not influence the jet's kinematic properties. These tracks must have $\pT>500~\MeV$ and $\vert\eta\vert<2.5$, and satisfy loose quality criteria and track-to-vertex matching requirements~\cite{PERF-2015-08} to reject fake tracks from the reconstruction and tracks from \pileup, respectively.

Electron candidates are reconstructed by matching tracks in the inner detector to topological energy clusters in the electromagnetic  calorimeter~\cite{EGAM-2018-01}. Muons are reconstructed using tracks in the muon spectrometer, matched to tracks in the inner detector if available~\cite{MUON-2018-03}. Electrons and muons are required to satisfy \textit{Loose} identification ($\sim$93\% and $\sim$98\% efficiency respectively) and \textit{Loose} isolation ($\sim$99\% and 98\% efficiency respectively) criteria, to have transverse momentum $\pT>18~\GeV$ and be reconstructed within $|\eta|<2.47$ ($e$) or $|\eta|<2.7$ ($\mu$) (but electrons within $1.37<|\eta|<1.52$ are excluded), and at least one must have $\pT>27~\GeV$.  An inner-detector track associated with an electron (muon) must have a transverse impact parameter significance $d_0/\sigma_{d_0}<5\,(3)$ relative to the beamline, and a longitudinal impact parameter $z_0$ satisfying $\vert \zzsth\vert < 0.5$~mm.  An overlap removal procedure resolves cases in which multiple electrons, muons or jets are reconstructed from the same detector signature.

Events are selected by a combination of single-electron or single-muon triggers for each data-taking period~\cite{TRIG-2018-05, TRIG-2018-01, TRIG-2019-03}, and the lepton reconstructed online by the trigger is required to be within $\Delta R=0.1$ of an offline reconstructed lepton. Events are required to have at least one reconstructed primary interaction vertex~\cite{ATL-PHYS-PUB-2015-026}. At least two same-flavour opposite-charge electrons or muons are required to pass this selection, and have an invariant mass compatible with the mass of the $Z$~boson: $81 <m_{\ell\ell}<101~\GeV$. If multiple same-flavour opposite-sign lepton pairs fulfil this requirement, the pair with an invariant mass closest to that of the $Z$~boson is chosen. A small fraction of the $Z\to\tau^+\tau^-$ decays can pass the selection if the two $\tau$-leptons decay to leptons of the same flavour. Higgs boson candidates are reconstructed from the lepton pair and jet system ($\ell\ell j$), which is required to have an invariant mass $m_{\ell\ell j}<250~\GeV$. If multiple jets satisfy these requirements, the jet with the highest \pT is selected. The jet is required to have at least two ghost-associated tracks.

\section{Background reweighting}

\label{sec:RW}

Jet substructure variables are crucial in order to improve the signal-to-background ratio. However, they are not modelled accurately enough, neither by \POWHEG nor \SHERPA for the $Z$\,+\,jets background. A binary classifier ($C(x)$) is known to be able to approximate the density ratio of two datasets when trained to distinguish between them, $p_1(x)/p_2(x) \approx C(x)/(1-C(x))$ \cite{cranmer2016approximatinglikelihoodratioscalibrated}. Evolving this approach, a neural network (NN) is trained to estimate the log-likelihood-ratio \cite{arxiv.1911.00405} of the total-background to data, with the goal of simultaneously improving the modelling of jet substructure and event kinematic variables. The exponential of the NN output is used as an additional weight for each MC event in order to equalize the two density functions. Events falling in the region $120<m_{\ell\ell j}<140~\GeV$, where most of the signal is expected, are excluded from the training and therefore only events in the $m_{\ell\ell j}$ sidebands [100-120] and [140-180]~\GeV are included. Any signal contamination outside this region is negligible. The NN is able to reweight events in the excluded region effectively. This is demonstrated by training another NN with a wider excluded region (115-145~\GeV) and comparing the distributions for every variable only for events lying in [115-120, 140-145]~\GeV region. No significant difference was observed.

The training variables consist of event kinematic variables (final-state invariant mass and transverse momentum, and dilepton and jet transverse momenta) and jet substructure variables. Seven dimensionless variables, constructed from the jet's ghost-associated tracks, are chosen to capitalize on the presence of a narrow jet or two-pronged substructure in the jet's track system: the number of tracks; the ratio of the \pT of the highest-\pT track to the \pT of the track system; the angular separation $\Delta R$ between the highest-\pT track and the jet axis; $N$-subjettiness $\tau_N$~\cite{Thaler:2010tr} (with $N=2$, exclusive-$k_t$-subjet axes with radius parameters of 0.2, and a jet-axis radius parameter of 0.4), which is an effective discriminating variable for tagging boosted objects; angularity ($\tilde{\tau}_{-2}$)~\cite{Almeida:2008yp}, which quantifies how the energy of the jet is distributed relative to the jet axis; and $U_1$ (with $\beta=0.7$)  and $M_2$ (with $\beta=0.3$), which are modified energy correlation functions~\cite{Moult:2016cvt} designed for quark--gluon discrimination and to target two-pronged substructure, respectively.

\begin{figure}[t!]
\centering
\begin{minipage}[b]{0.475\textwidth}
\centering
\includegraphics[width=\textwidth]{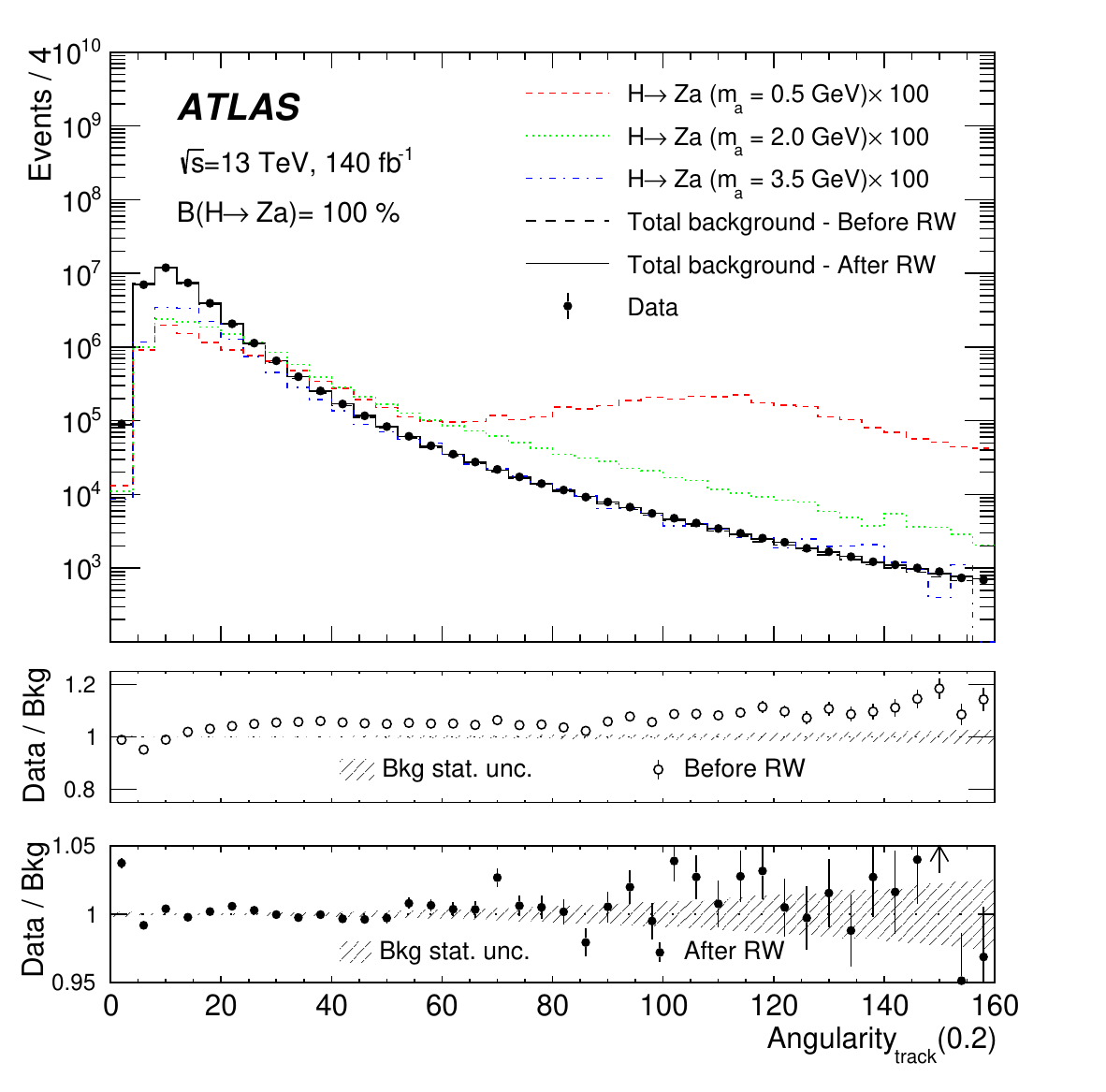}
\\[1ex]  %
(a)
\label{fig:minipage1}
\end{minipage}
\hfill
\begin{minipage}[b]{0.475\textwidth}
\centering
\includegraphics[width=\textwidth]{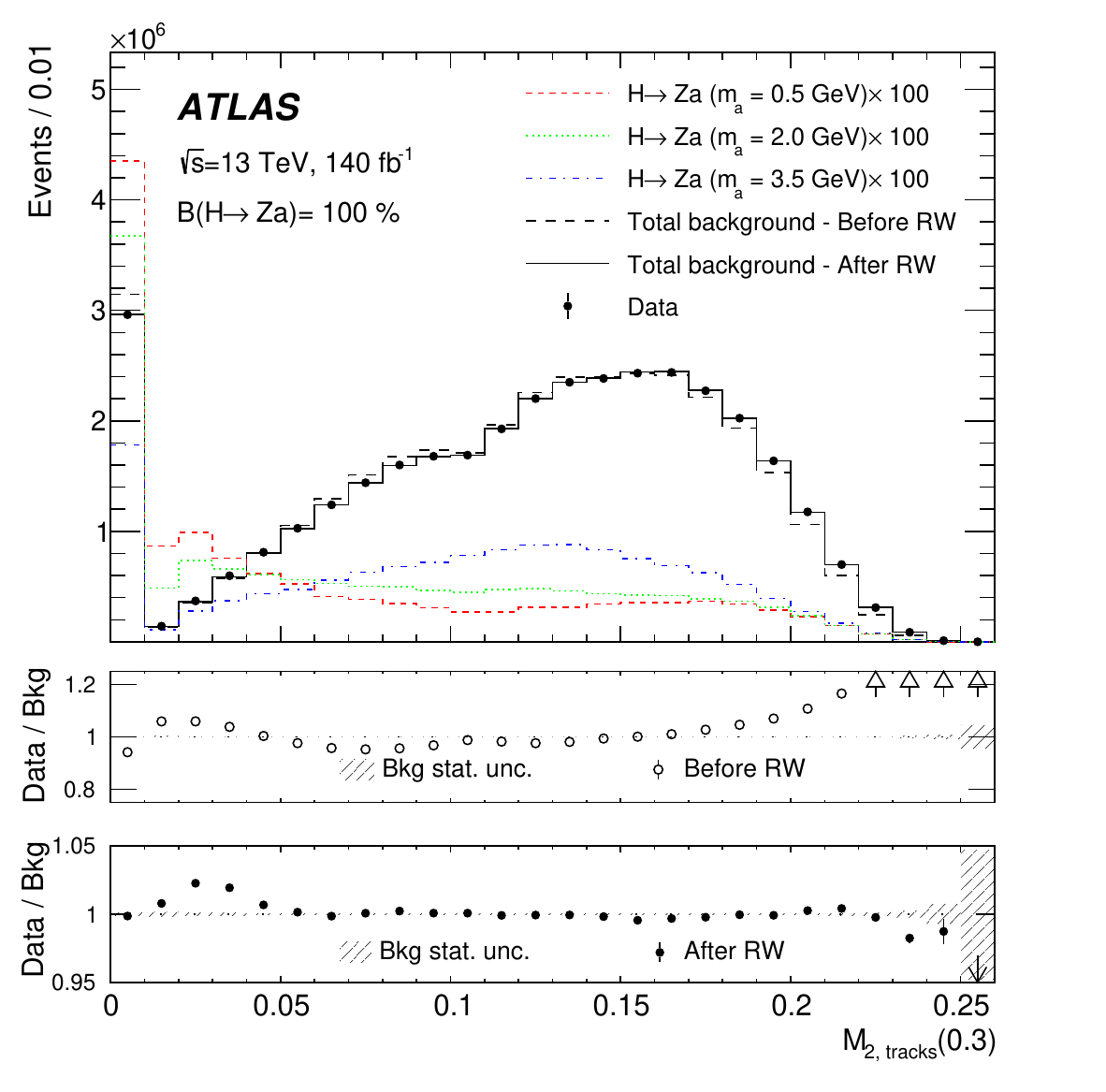}
\\[1ex]  %
(b)
\label{fig:minipage2}
\end{minipage}

\vspace{0.5cm} %

\begin{minipage}[b]{0.475\textwidth}
\centering
\includegraphics[width=\textwidth]{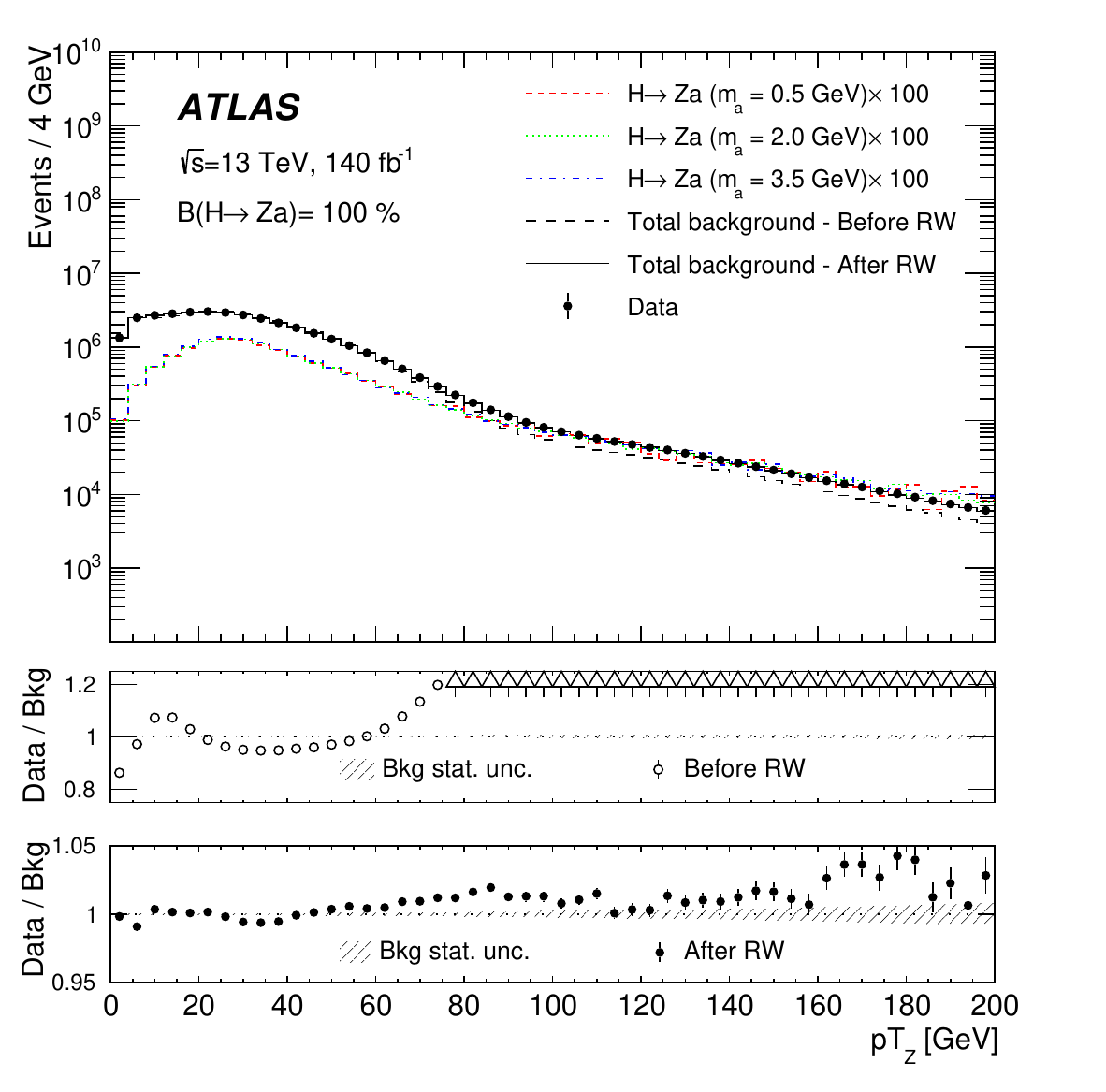}
\\[1ex]  %
(c)
\label{fig:minipage3}
\end{minipage}
\hfill
\begin{minipage}[b]{0.475\textwidth}
\centering
\includegraphics[width=\textwidth]{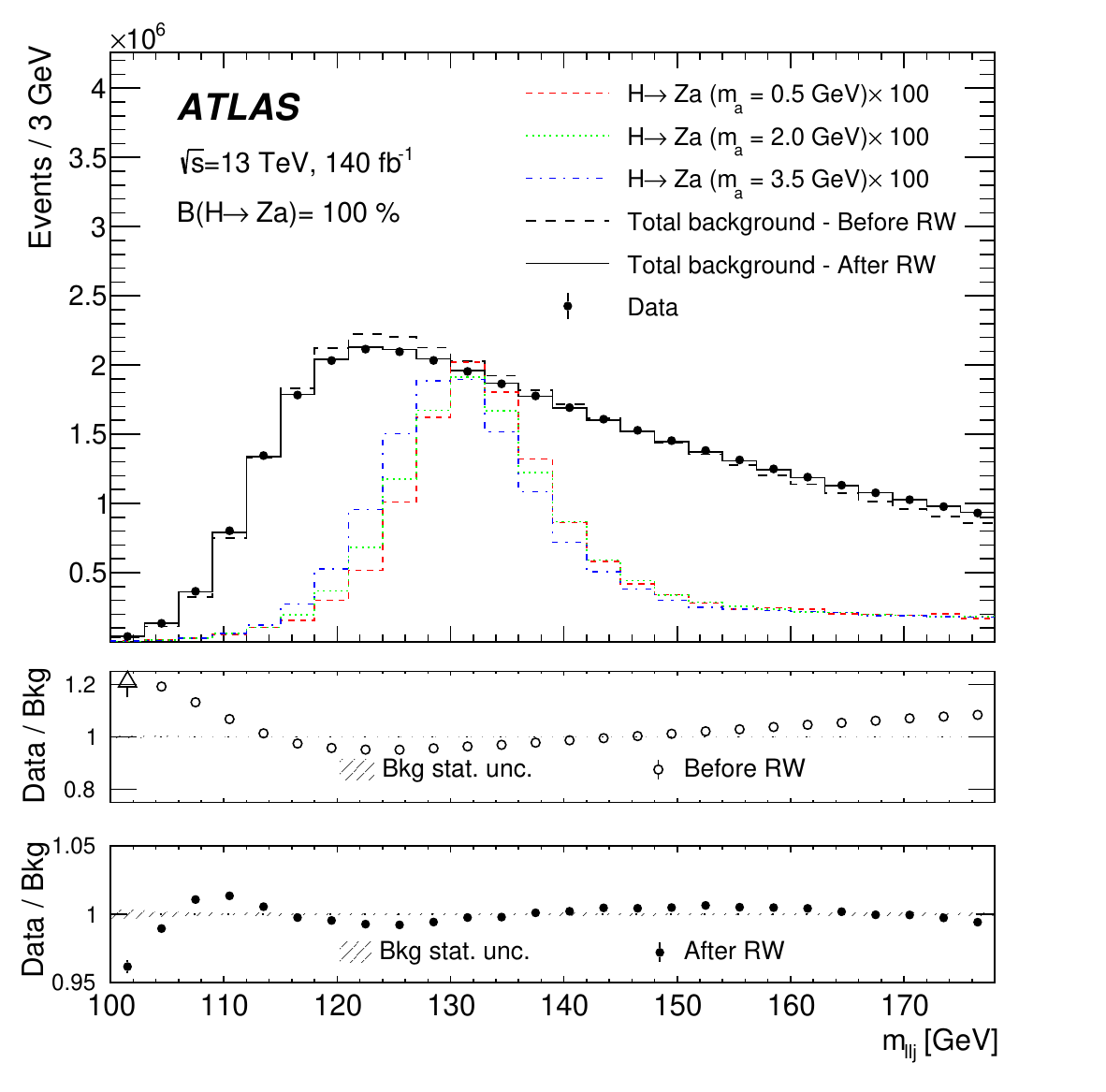}
\\[1ex]  %
(d)
\label{fig:minipage4}
\end{minipage}

\caption{The (a) angularity, (b) modified energy correlation function, (c) $Z$ boson transverse momentum, and (d) invariant mass of the lepton pair plus jet system, for data, background (pre- and post-reweighting) and three $H\rightarrow Za$ signal hypotheses (for $a\rightarrow q\bar{q}/gg$ inclusively). The two ratio panels show the data to background ratio before (top) and after (bottom) the reweighting. The lower panel uses a narrower $y$-axis range to capture the improvement in the background modelling. Events are required to pass the complete event selection but not the classification NN requirement. The background normalization is set equal to that of the data for events passing the preselection and being in the $m_{\ell\ell j}$ 100--180~\GeV region. The signal normalization assumes the SM Higgs boson inclusive production cross-section, $\mathcal{B}(H\to Za)=100\%$, and it is scaled up by a factor of 100. The error bars (hatched regions) represent the data (MC) sample's statistical uncertainty in the histograms and the ratio plots. Vertical arrows indicate data points that fall outside the displayed $y$-axis range.}
\label{fig:1}
\end{figure}

The reweighting NN is a simple feed-forward network designed and trained using Keras~\cite{chollet2015keras}. An exponential loss function for log-likelihood-ratio estimation ~\cite{arxiv.1911.00405} is used along with a hyperbolic tangent function ($8\tanh{x}$) which limits the maximum output of the last neuron. The ReLU function is chosen for neuron activation, as typically suggested for feed-forward NNs. For the loss minimizer, stochastic gradient descent \cite{ruder2016overview} was chosen over adaptive moment estimation (Adam~\cite{kingma2017adammethodstochasticoptimization}) as it provides more flexibility to tune its internal parameters. The number of layers (3), the number of neurons per layer (100), the learning rate (0.11), the decay rate (0.001), the momentum (0.975) and the batch size (200) are instead chosen via a hyperparameter optimization performed with a probabilistic Bayesian method. The goal of Bayesian optimization is to minimize the error function $f(x)$, which is a measure of the algorithm's mismodelling as a function of the hyperparameters. The set of hyperparameters is chosen with the use of Bayesian statistics for the prior and the observations. The optimization is implemented with the \textit{Hyperopt} library and the Tree-structured Parzen Estimator algorithm~\cite{arxiv.1209.5111}. This strategy reaches optimal performance significantly more quickly than other approaches such as random or grid searches. The training ends when the validation loss stops improving (with 10-epoch patience) and the epoch with the lowest loss value is chosen (epoch 50). The final-state invariant mass and three other variables used in the reweighting (RW) are shown in Figure~\ref{fig:1}.

\section{Signal--to--background discrimination}

\begin{figure}[t!]
\centering
\begin{minipage}[b]{0.475\textwidth}
\centering
\includegraphics[width=\textwidth]{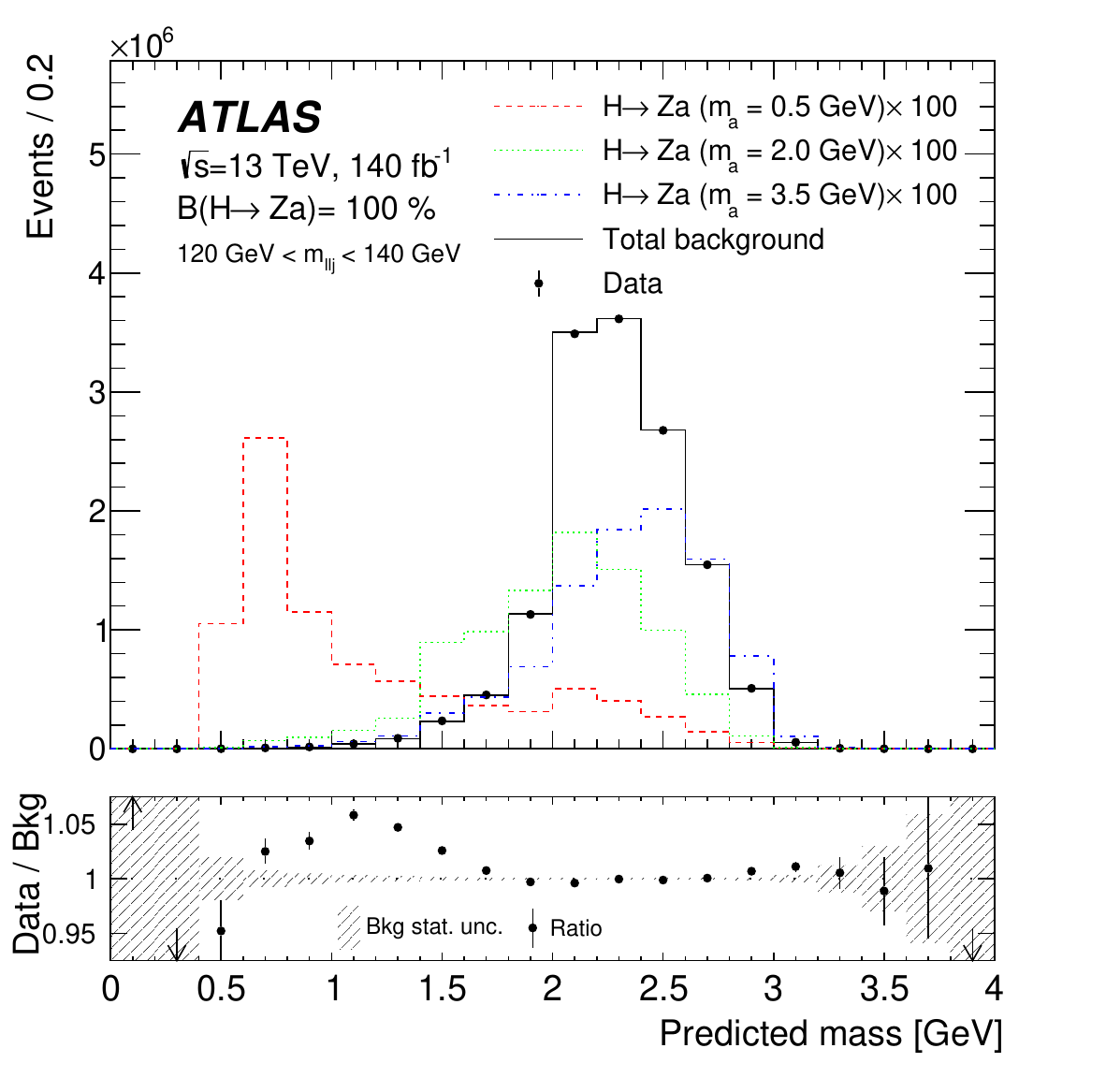}
\\[1ex]  %
(a)
\label{fig:minipage5}
\end{minipage}
\hfill
\begin{minipage}[b]{0.475\textwidth}
\centering
\includegraphics[width=\textwidth]{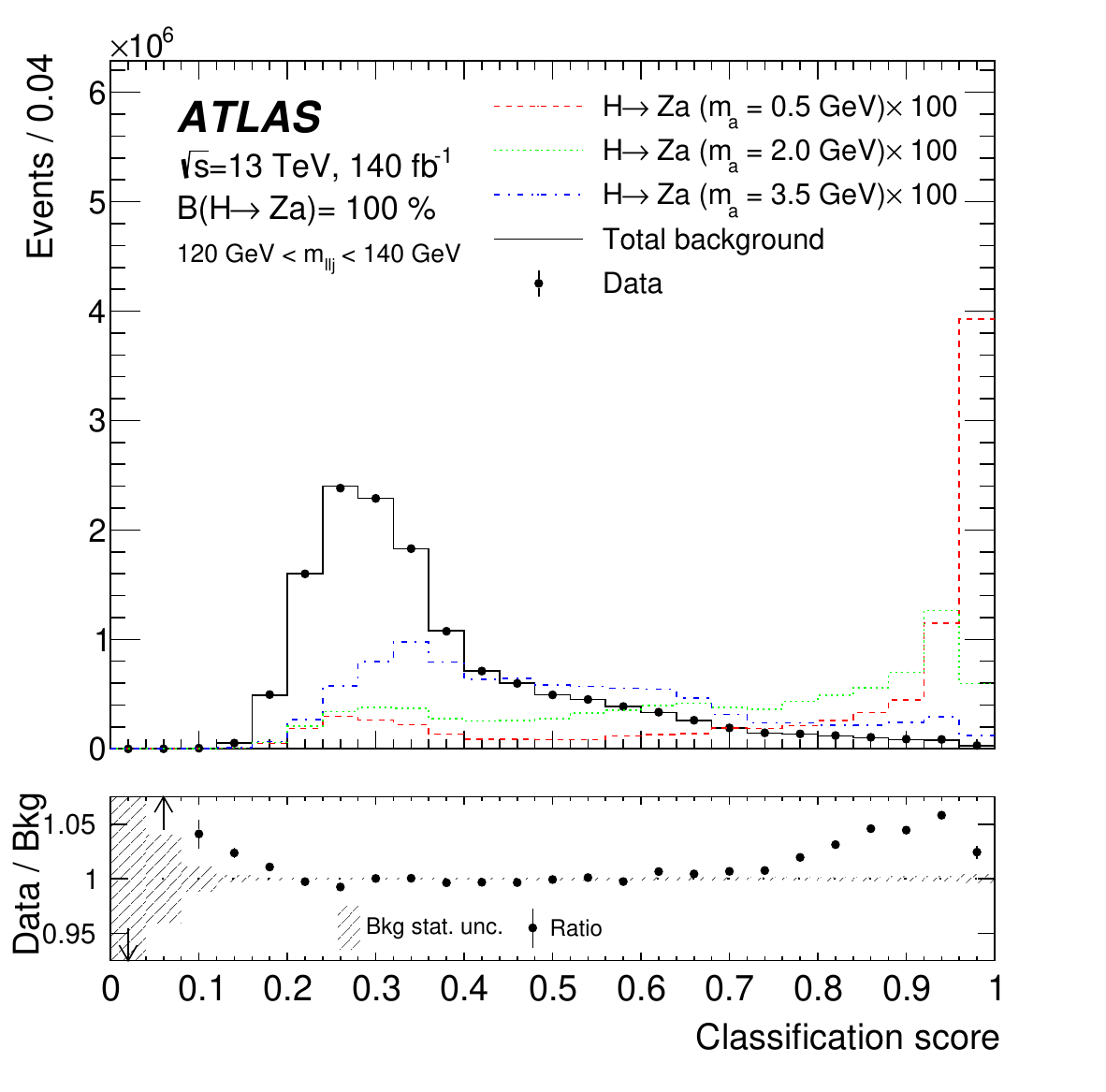}
\\[1ex]  %
(b)
\label{fig:minipage6}
\end{minipage}

\caption{The (a) regression and (b) classification NN output variables for data, reweighted background, and three $H\rightarrow Za$ signal hypotheses ($a\rightarrow q\bar{q}/gg$ inclusively). Events are required to pass the complete event selection, including a $120<m_{\ell\ell j}<140$~\GeV requirement, but not the classification NN output variable requirement. The background normalization is set equal to that of the data for events passing the preselection and being in the $m_{\ell\ell j}$ 100--180~\GeV region. The signal normalization assumes the SM Higgs boson inclusive production cross-section, $\mathcal{B}(H\to Za)=100\%$, and it is scaled up by a factor of 100. The error bars (hatched regions) represent the data (MC) sample's statistical uncertainty in the histograms and the ratio plots. Vertical arrows indicate data points that fall outside the displayed $y$-axis range. }
\label{fig:2}
\end{figure}

After the initial event selection, two NNs are used to further suppress the background. First, a \emph{regression} NN is used to estimate the mass of the $a$-resonance, and second, a \emph{classification} NN is used to discriminate between signal and background. The reconstructed jet mass resolution and scale for such low-energy jets are not enough to inform the classifier about the different mass hypotheses (the jet mass distributions are similar for all of them). Therefore, a regression NN is trained with the seven jet substructure variables to estimate the mass of the resonance (two of the variables are shown in Figures~\ref{fig:1}(a) and~\ref{fig:1}(b)).  Its output in Figure~\ref{fig:2}(a) differentiates between the various signal hypotheses significantly better than the jet mass. It is then provided along with the same seven variables to the classification NN. The regression-NN strategy is preferable to other commonly used approaches (e.g.\ a parameterized NN using the generator-level resonance mass) as it allows the analysis to be conducted with a single classification NN and a single background model, without a significant decrease in performance.

The regression NN is trained on nine different signal-hypothesis samples ranging from 0.5 to 4~\GeV in steps of 0.5~\GeV. The classification NN is trained with all the signal hypotheses against the total background. Both the quark and gluon decay modes are included and their relative contributions come from the MSSM scenario mentioned in Sec.~\ref{sec:DataAndMC}. The fraction of $gg$ decays varies from 97\% to 6\% for masses of 0.5 to 4~\GeV and the $q\bar{q}$ decay fraction varies correspondingly from 0\% to 56\%. The hyperparameter optimization follows the same methodology used for the reweighting, encompassing typically suggested functions, self-optimizing algorithms and Bayesian optimization. Both the NN performance and the level of MC--data agreement ($\chi^2$ test) for the output distributions are taken into account to select the best set. The optimized hyperparameter values for the regression\,/\,classification NN are as follows:\ number of layers (4\,/\,3), number of neurons per layer (35\,/\,10), neuron activation function (ReLU\,/\,ReLU), output function (linear\,/\,sigmoid), loss function (Huber\,/\,binary cross-entropy), loss minimizer (Adam\,/\,Adam), batch size (160\,/\,100), regularization (R2, 10$^{-11}$\,/\,R2, 10$^{-10}$) and epochs (50\,/\,29). Figure~\ref{fig:2} shows the regression and classification NN output variables for the reweighted background, the data and three signal hypotheses.

Although the NNs are trained with a mixture of $a\rightarrow gg$ and $a\rightarrow q\bar{q}$ decays (with different branching fractions), their performance is generally similar for the two decay modes. %
Their performance is poorer for the higher masses, where $a$-resonance decays tend to have higher track multiplicity, making the hadronic decay appear similar to background jets. This is reflected in the distributions of jet substructure training variables, which are indeed very similar for higher-mass signals and background in Figures~\ref{fig:1}(a) and~\ref{fig:1}(b). The $a\to c\bar c$ decay mode becomes dominant in the 4~\GeV sample, leading to quite different kinematics and much poorer NN performance. Consequently, the 4~\GeV $a\to q\bar q$ case is omitted from the final results.

A single threshold requirement is placed on the classification NN's output variable to reject background events for all signal hypotheses. The threshold value is chosen to be 0.93, which rejects 99.3\% of the background for a signal efficiency of more than 30\% for $m_a = 0.5$~\GeV, $\sim$10\% for $m_a = 2$~\GeV, and $\sim$3\% for $m_a = 3.5$~\GeV. Although this threshold value is not optimal for the higher masses, it allows a single background model to be used. Events passing the classification NN selection populate the signal region (SR). %

\begin{figure}[t!]
\centering
\begin{minipage}[b]{0.475\textwidth}
\centering
\includegraphics[width=\textwidth]{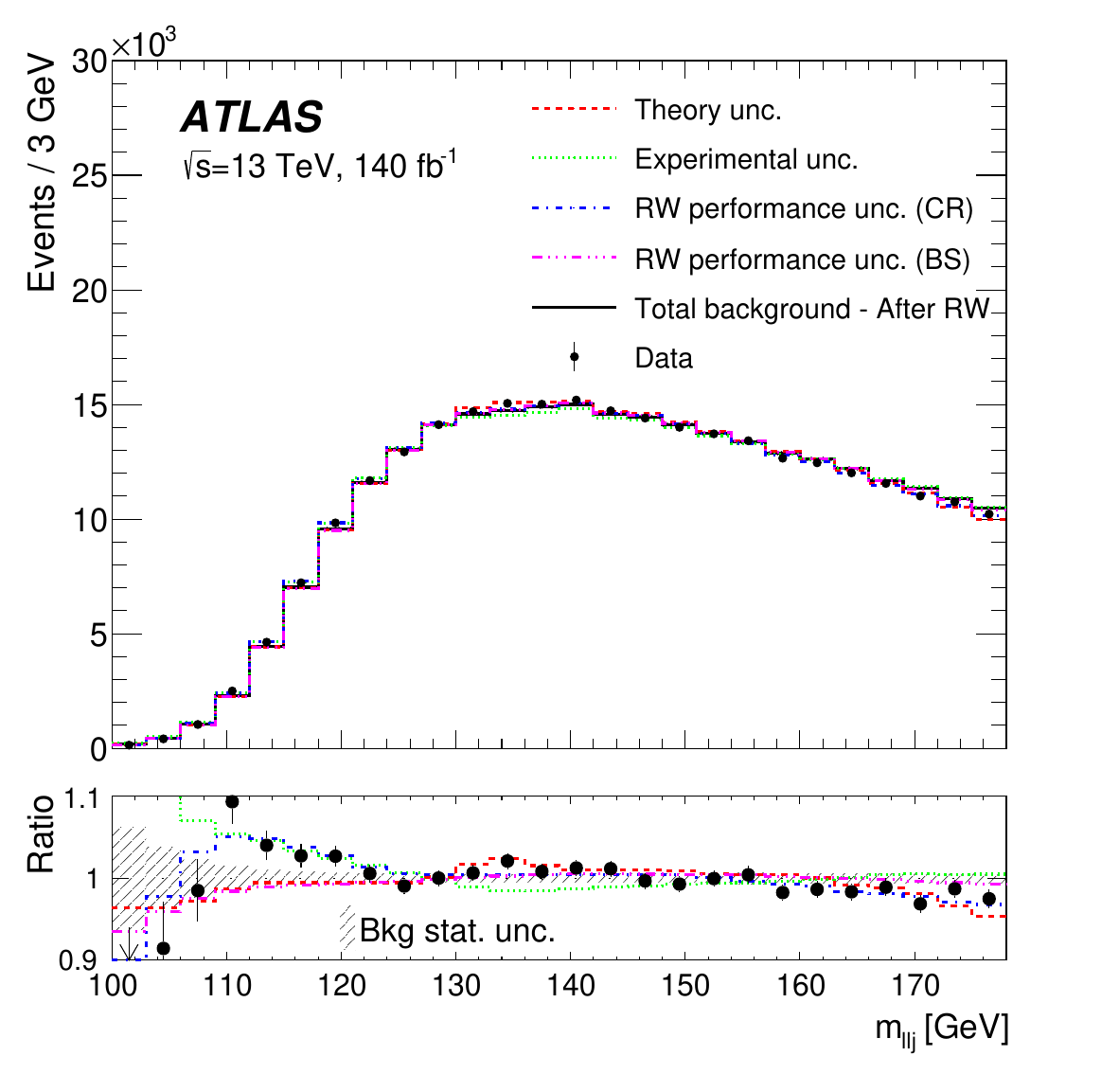}
\\[1ex]  %
(a)
\label{fig:minipage7}
\end{minipage}
\hfill
\begin{minipage}[b]{0.475\textwidth}
\centering
\includegraphics[width=\textwidth]{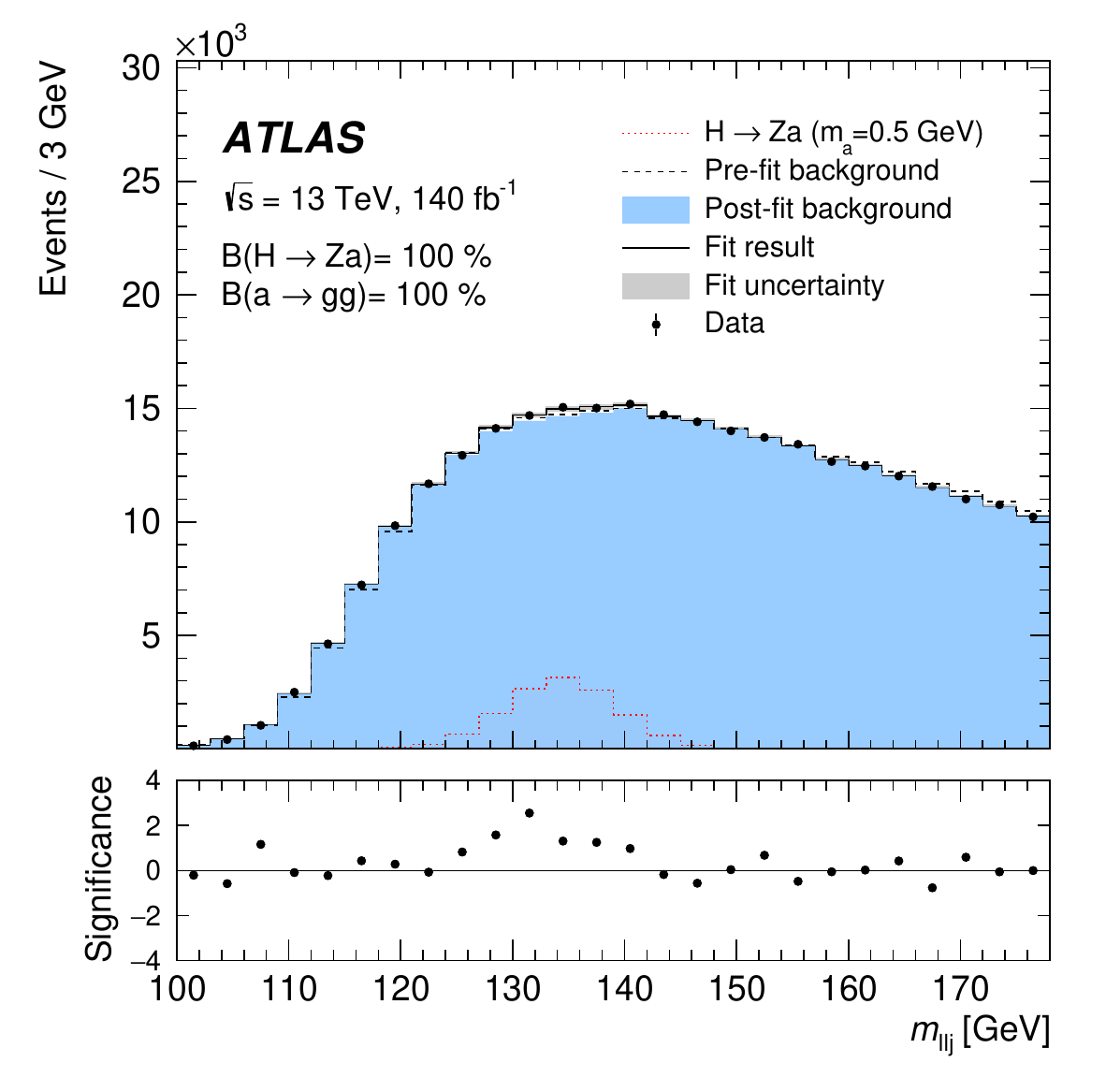}
\\[1ex]  %
(b)
\label{fig:minipage8}
\end{minipage}

\caption{Invariant mass of the lepton pair plus jet system for data, reweighted background and its systematic uncertainties, and a signal hypothesis. Events are required to pass the complete event selection (preselection and classification NN requirement). In (a) the pre-fit background distribution is shown along with its systematic uncertainties (only "up" variations are shown).
The background normalization is set equal to that of the data for events passing the preselection and being in the 100--180~\GeV and  Classification Score $>0.93$ region. The experimental uncertainty combines all the relevant jet uncertainties assuming they are completely uncorrelated. The lower panel shows the ratio of the data and the various uncertainties to the pre-fit background prediction. In (b) the background shape and normalization come from the fit to the data. The signal includes only $gg$ decays, its shape comes from the fit, and it is normalized to $\mathcal{B}(H\rightarrow Za)=100\%$. The lower panels shows the ratio of the data and various uncertainties to the pre-fit background prediction and (b) the per-bin signal significance with respect to the post-fit background, $(N_\text{data}-N_\text{bkg})\,\text{\Large{/}}\sqrt{\sigma_\text{data}^2+\sigma_\text{bkg}^2}$. The post-fit background uncertainty is calculated from a fit under the background-only hypothesis. The error bars represent the data sample's statistical uncertainty and the grey bands represent the background model statistical uncertainty, (a) before the fit and (b) after the fit. Vertical arrows indicate data points that fall outside the displayed $y$-axis range.}
\label{fig:3}
\end{figure}

\section{Statistical model}

A measure of the signal strength is extracted for each given signal hypothesis using a binned maximum-likelihood fit~\cite{Cowan:2010js-witherratum} to the distribution of the final-state invariant mass $m_{\ell\ell j}$ in the range 100--178~\GeV, as shown in Figure~\ref{fig:3}(b). The signal strength is translated to $\mathcal{B}(H\to Za)$ assuming $\mathcal{B}(a\rightarrow gg)=100\%$ or $\mathcal{B}(a\rightarrow q\bar{q})=100\%$. Systematic uncertainties are included in the fit as nuisance parameters which modify the signal and background model and they are implemented as Gaussian constraints on the nominal values. Yet the total background normalization is included as a free floating uncertainty.

The classification NN selection removes the event signal events on the right tail, which come from a mis-selected jet as $a$ candidate. Therefore, the signal model is produced from a simple Gaussian fit with mean $\mu$ and width $\sigma$ of the MC-simulated $m_{\ell\ell j}$ distribution. %
Different $(\mu, \sigma)$ values are obtained for each mass point and decay mode ($a\rightarrow gg/q\bar{q}$). The signal's mean ($\mu$) shifts from 125~\GeV for $m_a=4$~\GeV to 131.5~\GeV for $m_a=0.5$~\GeV. This mass dependence is expected, as higher masses result in less collimated jets, causing some energy to fall outside the jet cone. Additionally, the jet calibration is not optimized for such low energies, contributing to the shift in the mean. A mean value around 1~\GeV higher is observed for $a\rightarrow gg$ decays. The signal's width ($\sigma$) of 5.0--5.5~\GeV varies little with the decay mode or $m_a$ value. For the background the reweighted MC shape is used.

Systematic uncertainties are the dominant source of uncertainty for this analysis. For the signal, both the theory and experimental uncertainties are considered. They can modify the predicted signal yield in the SR, shift the mass peak ($\mu$), and alter the width of the signal peak ($\sigma$).
The experimental uncertainties come from the jet reconstruction, and the lepton reconstruction, identification, isolation and track-to-vertex matching, as well as from the \pileup distribution and the trigger efficiency. The jet uncertainties (energy scale and resolution) are found to have the largest effect. The experimental uncertainties have minor impact on the signal yield but they affect the mean ($\mu$) and width ($\sigma$),  resulting in a combined uncertainty of about 1.0~\GeV for both ($\Delta\mu=1.1$~\GeV and $\Delta\sigma=0.7$~\GeV). Theory uncertainties in parton showering and hadronization are estimated by comparing the nominal \PYTHIA MC signal samples with alternative \HERWIG samples. The \HERWIG samples are produced separately for $a\rightarrow gg$ and $a\rightarrow q\bar{q}$, but only for masses greater than 2~\GeV due to \HERWIG technical specifications. The uncertainties are extrapolated to the lower masses. They affect the Gaussian mean ($\mu$), resulting in $\Delta\mu=0.5{-}2.5$~\GeV, but their effect on the Gaussian width is minor. In general, the effect on $\mu$ tends to decrease for higher masses and it is $\sim$1~\GeV larger for $gg$ decays than for $q\bar{q}$ decays. More importantly, the theory uncertainties affect the classification NN's efficiency, leading to an uncertainty in the expected number of signal events of about 60\% for $a\rightarrow gg$ and about 25\% for $a\rightarrow q\bar{q}$. This is the most significant uncertainty for this analysis, and it is due to \PYTHIA generally decaying the $a$-resonance with lower multiplicity than \HERWIG. The NN is particularly effective in identifying two-track jets, and these are more common in the \PYTHIA samples. For $\eta_c$ and $J/\psi$, uncertainties from the $m_a=3.0$~\GeV model are used. In addition, a total Higgs boson production cross-section uncertainty of ${^{\,+5.6\%}_{-7.4\%}}$ is included~\cite{deFlorian:2016spz}.

Five sources of uncertainty incorporate variations in the reweighting performance, as well as experimental and theoretical uncertainties in the background modelling. All of them are included in the fit as alternative shapes for the background distribution. For the NN performance, two uncertainties are considered. The first is estimated from a control region (CR) that includes the same number of background events as the SR. It is defined by changing the classification NN requirement from $>$0.93 (SR) to 0.883$-$0.93. The data-MC discrepancy in the two regions was found very similar, motivating the addition of this uncertainty. %
For any reasonable $H\to Za$ branching fraction ($<$20\%), the resulting signal contamination has a negligible effect in the shape of the systematic distribution produced in the CR. The ratio of data events to MC events in this region is used to derive a shape systematic uncertainty. The second is estimated using a bootstrapping (BS) method. Ten replicas of the original MC and data event samples are created by weighting each event with a random Poisson($\mu=1$) number. For each replica a reweighting NN is trained, and a systematic uncertainty distribution is built from the bin-by-bin standard deviations in the SR. The main sources of experimental uncertainty are the jet energy scale and resolution. MC samples are produced with corresponding scale and resolution variations, their predictions are reweighted using the nominal model, and the resulting shape differences are used. Other experimental uncertainties, such as tracking uncertainties, are found to be insignificant. The effect of theory uncertainties in the matrix element generator, in the parton showering and hadronization are estimated
by comparing the nominal $Z$\,+\,jets background model from \POWHEG+\PYTHIA with an alternative from \SHERPA. A second reweighting NN is trained using this sample, with the same input variables and training procedure as for the nominal \POWHEG sample. The ratio of the distributions in the SR is used as an additional systematic uncertainty. The MC statistical uncertainty is also taken into account. The final background model with its uncertainties is shown in Figure~\ref{fig:3}(a).

Finally, a luminosity uncertainty is included. The uncertainty in the combined 2015--2018 integrated luminosity is 0.83\%~\cite{DAPR-2021-01}, obtained using the LUCID-2 detector~\cite{LUCID2} for the primary luminosity measurements, complemented by measurements using the inner detector and calorimeters.

\section{Results and interpretations}

\begin{figure}[t!]
\centering
\begin{minipage}[b]{0.475\textwidth}
\centering
\includegraphics[width=\textwidth]{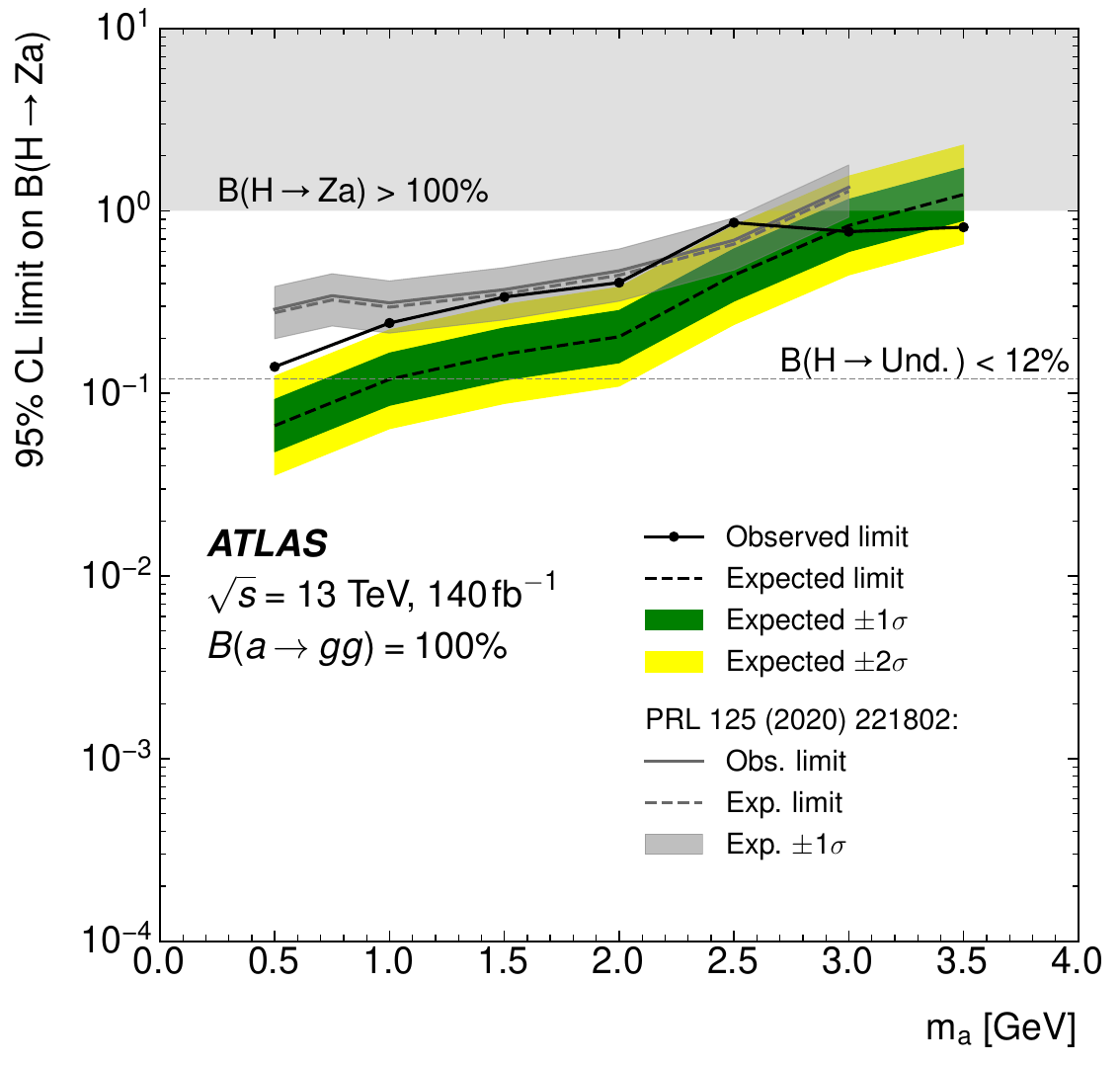}
\\[1ex]  %
(a)
\label{fig:minipage9}
\end{minipage}
\hfill
\begin{minipage}[b]{0.475\textwidth}
\centering
\includegraphics[width=\textwidth]{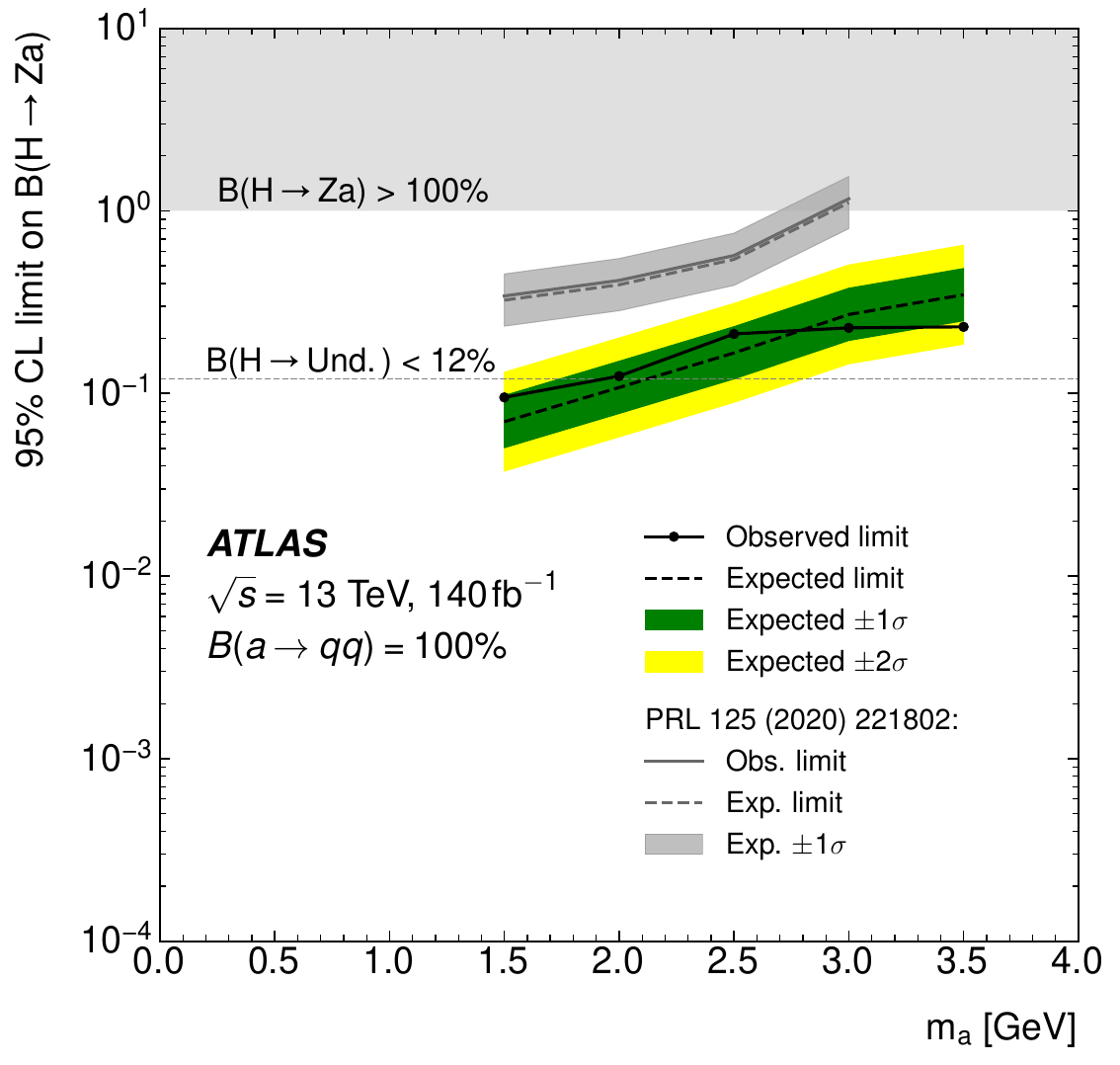}
\\[1ex]  %
(b)
\label{fig:minipage10}
\end{minipage}

\caption{Observed 95\% CL upper limits on $\mathcal{B}(H\to Za)$ for (a) $\mathcal{B}(a\rightarrow gg)=100\%$ and (b) $\mathcal{B}(a\rightarrow q\bar{q})=100\%$, and the expectation under the background-only hypothesis together with its $\pm1\sigma$ and $\pm2\sigma$ intervals. The weaker limits from the previous version of the analysis ~\cite{HDBS-2018-37} are also shown. Linear interpolation is used to set limits between the mass points for which MC signal samples were generated.The horizontal dashed lines indicate the model-independent bounds on Higgs decays to undetected particles as set in ~\cite{HIGG-2021-23}.}
\label{fig:4}
\end{figure}

Signal hypotheses leading to an expected $m_{\ell\ell j}$ distribution with mean larger than 130~\GeV (corresponding to $m_a=0.5{-}2.5$~\GeV and $a\to gg$) accommodate an excess of events observed in data around 135~\GeV as reported in Figure~\ref{fig:3}(b). Conversely, if the signal mean is lower, the data are more compatible with the background only hypothesis. The largest local significance is observed for $m_a=0.5$ GeV in the  $a\to gg$ scenario and it is $\sim$ 1.5$\sigma$, indicating that the data are compatible with the SM scenario. Upper limits at 95\% CL are set (assuming SM Higgs production cross section) on $\mathcal{B}(H\to Za)$ for the various signal hypotheses, using the profile-likelihood test statistic~\cite{Cowan:2010js-witherratum} and the $CL_S$ technique (asymptotic formulae) ~\cite{Read:2002hq}. The observed and expected upper limits for $a\rightarrow gg$ and $a\rightarrow q\bar{q}$ are shown in Figure~\ref{fig:4}, in comparison to the limits from Ref. ~\cite{HDBS-2018-37}. For $a\rightarrow q\bar{q}$, only decays to the heaviest possible quarks are considered. The slight excess around 135~\GeV mentioned above, causes the observed limits to be higher than the expected limits for cases where the signal mean is greater than $\sim$130~\GeV. The expected limits are generally more restrictive for $a\rightarrow q\bar{q}$ because of the lower parton shower and hadronization uncertainty. The upper limit values range from $\sim$10\% for 0.5 \GeV to $>$80\% for 3.5 \GeV signal for the $gg$ scenario, and from $<$10\% for 1.5 GeV to $\sim$20\% for 3.5 \GeV for the $q\bar{q}$ scenario.

Limits are also set on $H\to Z\,\eta_c$ and $H\to Z\,J/\psi$ decays; however, the analysis has limited sensitivity for these cases and cannot exclude any physical branching fraction. It excludes $\mathcal{B}(H\to Z\,\eta_c)>1.2$ and $\mathcal{B}(H\to Z\,J/\psi)>1.4$, corresponding to a $\sim30\%$ improvement with respect to Ref.~\cite{HDBS-2018-37}. Finally, ALP decay rates can be obtained by assuming the effective Wilson coefficients (determining the coupling strength to various particles) are equal to 1~\cite{Bauer:2017ris}. %
For ALP masses of 0.5--1~\GeV the $3\pi$ final state is one of the leading decay modes. Only the $a\to\pi^+\pi^-\pi^0$ decay is considered, as the analysis is not sensitive to $a\to3\pi^0$, due to the lack of inner-detector tracks. For $\mathcal{B}(a\to\pi^+\pi^-\pi^0)=0.10$~(0.20) and $m_a=0.5$~\GeV~(1~\GeV), the upper limit on $\mathcal{B}(H\to Za)$ is 0.45~(0.81). This can be used to exclude values of the effective coupling $C^\text{eff}_{ZH}/\Lambda$ of the candidate axion to the Higgs and $Z$ bosons. At 95\% CL, $C^\text{eff}_{ZH}/\Lambda<2.0~\TeV^{-1}$ for $m_a=0.5$~\GeV and $C^\text{eff}_{ZH}/\Lambda<0.89~\TeV^{-1}$ for $m_a=1$~\GeV.

\FloatBarrier

\section{Conclusions}
\label{sec:conclusion}
A search for Higgs boson decays into a $Z$~boson and a light resonance of mass 0.5--3.5~\GeV is performed using 140~fb$^{-1}$ of proton—proton collisions at $\sqrt s=13$~\TeV recorded by the ATLAS detector at the LHC. The search considers leptonic decays of the $Z$~boson and hadronic decays of the light resonance that are collimated into a single small-radius jet. Separate neural networks are used to improve the modelling of the simulated background, and to discriminate signal from background. A binned profile-likelihood fit is performed on the invariant mass of the dilepton-plus-jet system. No significant excess above the Standard Model background prediction is found and upper limits at 95\% confidence level are set on the branching fraction for $H \to Za$, where $a$ is a light pseudoscalar, as a function of its mass $m_a$, separately for the $a \to gg$ and $a \to q\bar{q}$ scenarios. Compared to a previous ATLAS search using the same dataset, the innovative techniques used in this analysis allow to improve the branching fraction limits by a factor of up to two (three) in the case of $a \to gg$ ($a \to q\bar{q}$). Upper limits are also set on the branching fractions for $H\to Z\,\eta_c$ and $H\to Z\,J/\psi$, and on the effective coupling of an axion-like particle to the Higgs boson and $Z$~boson ($C^\text{eff}_{ZH}/\Lambda$).

\section*{Acknowledgements}
%

%

%
%

%
%

We thank CERN for the very successful operation of the LHC and its injectors, as well as the support staff at
CERN and at our institutions worldwide without whom ATLAS could not be operated efficiently.

The crucial computing support from all WLCG partners is acknowledged gratefully, in particular from CERN, the ATLAS Tier-1 facilities at TRIUMF/SFU (Canada), NDGF (Denmark, Norway, Sweden), CC-IN2P3 (France), KIT/GridKA (Germany), INFN-CNAF (Italy), NL-T1 (Netherlands), PIC (Spain), RAL (UK) and BNL (USA), the Tier-2 facilities worldwide and large non-WLCG resource providers. Major contributors of computing resources are listed in Ref.~\cite{ATL-SOFT-PUB-2025-001}.

We gratefully acknowledge the support of ANPCyT, Argentina; YerPhI, Armenia; ARC, Australia; BMWFW and FWF, Austria; ANAS, Azerbaijan; CNPq and FAPESP, Brazil; NSERC, NRC and CFI, Canada; CERN; ANID, Chile; CAS, MOST and NSFC, China; Minciencias, Colombia; MEYS CR, Czech Republic; DNRF and DNSRC, Denmark; IN2P3-CNRS and CEA-DRF/IRFU, France; SRNSFG, Georgia; BMBF, HGF and MPG, Germany; GSRI, Greece; RGC and Hong Kong SAR, China; ICHEP and Academy of Sciences and Humanities, Israel; INFN, Italy; MEXT and JSPS, Japan; CNRST, Morocco; NWO, Netherlands; RCN, Norway; MNiSW, Poland; FCT, Portugal; MNE/IFA, Romania; MSTDI, Serbia; MSSR, Slovakia; ARIS and MVZI, Slovenia; DSI/NRF, South Africa; MICIU/AEI, Spain; SRC and Wallenberg Foundation, Sweden; SERI, SNSF and Cantons of Bern and Geneva, Switzerland; NSTC, Taipei; TENMAK, T\"urkiye; STFC/UKRI, United Kingdom; DOE and NSF, United States of America.

Individual groups and members have received support from BCKDF, CANARIE, CRC and DRAC, Canada; CERN-CZ, FORTE and PRIMUS, Czech Republic; COST, ERC, ERDF, Horizon 2020, ICSC-NextGenerationEU and Marie Sk{\l}odowska-Curie Actions, European Union; Investissements d'Avenir Labex, Investissements d'Avenir Idex and ANR, France; DFG and AvH Foundation, Germany; Herakleitos, Thales and Aristeia programmes co-financed by EU-ESF and the Greek NSRF, Greece; BSF-NSF and MINERVA, Israel; NCN and NAWA, Poland; La Caixa Banking Foundation, CERCA Programme Generalitat de Catalunya and PROMETEO and GenT Programmes Generalitat Valenciana, Spain; G\"{o}ran Gustafssons Stiftelse, Sweden; The Royal Society and Leverhulme Trust, United Kingdom.

In addition, individual members wish to acknowledge support from Armenia: Yerevan Physics Institute (FAPERJ); CERN: European Organization for Nuclear Research (CERN PJAS); Chile: Agencia Nacional de Investigaci\'on y Desarrollo (FONDECYT 1230812, FONDECYT 1230987, FONDECYT 1240864); China: Chinese Ministry of Science and Technology (MOST-2023YFA1605700, MOST-2023YFA1609300), National Natural Science Foundation of China (NSFC - 12175119, NSFC 12275265, NSFC-12075060); Czech Republic: Czech Science Foundation (GACR - 24-11373S), Ministry of Education Youth and Sports (ERC-CZ-LL2327, FORTE CZ.02.01.01/00/22\_008/0004632), PRIMUS Research Programme (PRIMUS/21/SCI/017); EU: H2020 European Research Council (ERC - 101002463); European Union: European Research Council (ERC - 948254, ERC 101089007), Horizon 2020 Framework Programme (MUCCA - CHIST-ERA-19-XAI-00), European Union, Future Artificial Intelligence Research (FAIR-NextGenerationEU PE00000013), Italian Center for High Performance Computing, Big Data and Quantum Computing (ICSC, NextGenerationEU); France: Agence Nationale de la Recherche (ANR-20-CE31-0013, ANR-21-CE31-0013, ANR-21-CE31-0022, ANR-22-EDIR-0002); Germany: Baden-Württemberg Stiftung (BW Stiftung-Postdoc Eliteprogramme), Deutsche Forschungsgemeinschaft (DFG - 469666862, DFG - CR 312/5-2); Italy: Istituto Nazionale di Fisica Nucleare (ICSC, NextGenerationEU), Ministero dell'Università e della Ricerca (NextGenEU I53D23000820006 M4C2.1.1); Japan: Japan Society for the Promotion of Science (JSPS KAKENHI JP22H01227, JSPS KAKENHI JP22H04944, JSPS KAKENHI JP22KK0227, JSPS KAKENHI JP23KK0245); Netherlands: Netherlands Organisation for Scientific Research (NWO Veni 2020 - VI.Veni.202.179); Norway: Research Council of Norway (RCN-314472); Poland: Ministry of Science and Higher Education (IDUB AGH, POB8, D4 no 9722), Polish National Agency for Academic Exchange (PPN/PPO/2020/1/00002/U/00001), Polish National Science Centre (NCN 2021/42/E/ST2/00350, NCN OPUS 2023/51/B/ST2/02507, NCN OPUS nr 2022/47/B/ST2/03059, NCN UMO-2019/34/E/ST2/00393, NCN \& H2020 MSCA 945339, UMO-2020/37/B/ST2/01043, UMO-2021/40/C/ST2/00187, UMO-2022/47/O/ST2/00148, UMO-2023/49/B/ST2/04085, UMO-2023/51/B/ST2/00920); Portugal: Foundation for Science and Technology (FCT); Slovenia: Slovenian Research Agency (ARIS grant J1-3010); Spain: Generalitat Valenciana (Artemisa, FEDER, IDIFEDER/2018/048), Ministry of Science and Innovation (MCIN \& NextGenEU PCI2022-135018-2, MICIN \& FEDER PID2021-125273NB, RYC2019-028510-I, RYC2020-030254-I, RYC2021-031273-I, RYC2022-038164-I); Sweden: Carl Trygger Foundation (Carl Trygger Foundation CTS 22:2312), Swedish Research Council (Swedish Research Council 2023-04654, VR 2018-00482, VR 2021-03651, VR 2022-03845, VR 2022-04683, VR 2023-03403), Knut and Alice Wallenberg Foundation (KAW 2018.0458, KAW 2019.0447, KAW 2022.0358); Switzerland: Swiss National Science Foundation (SNSF - PCEFP2\_194658); United Kingdom: Leverhulme Trust (Leverhulme Trust RPG-2020-004), Royal Society (NIF-R1-231091); United States of America: U.S. Department of Energy (ECA DE-AC02-76SF00515), Neubauer Family Foundation.

%
%


%
\clearpage
\printbibliography
\clearpage
 
\begin{flushleft}
\hypersetup{urlcolor=black}
{\Large The ATLAS Collaboration}

\bigskip

\AtlasOrcid[0000-0002-6665-4934]{G.~Aad}$^\textrm{\scriptsize 105}$,
\AtlasOrcid[0000-0001-7616-1554]{E.~Aakvaag}$^\textrm{\scriptsize 17}$,
\AtlasOrcid[0000-0002-5888-2734]{B.~Abbott}$^\textrm{\scriptsize 124}$,
\AtlasOrcid[0000-0002-0287-5869]{S.~Abdelhameed}$^\textrm{\scriptsize 120a}$,
\AtlasOrcid[0000-0002-1002-1652]{K.~Abeling}$^\textrm{\scriptsize 57}$,
\AtlasOrcid[0000-0001-5763-2760]{N.J.~Abicht}$^\textrm{\scriptsize 51}$,
\AtlasOrcid[0000-0002-8496-9294]{S.H.~Abidi}$^\textrm{\scriptsize 30}$,
\AtlasOrcid[0009-0003-6578-220X]{M.~Aboelela}$^\textrm{\scriptsize 46}$,
\AtlasOrcid[0000-0002-9987-2292]{A.~Aboulhorma}$^\textrm{\scriptsize 36e}$,
\AtlasOrcid[0000-0001-5329-6640]{H.~Abramowicz}$^\textrm{\scriptsize 156}$
% ...
\bigskip
\\

% ...
$^{17}$Department for Physics and Technology, University of Bergen, Bergen; Norway.\\
% ...
$^{30}$Physics Department, Brookhaven National Laboratory, Upton NY; United States of America.\\
% ...
$^{36}$$^{(a)}$Facult\'e des Sciences Ain Chock, Universit\'e Hassan II de Casablanca;$^{(b)}$Facult\'{e} des Sciences, Universit\'{e} Ibn-Tofail, K\'{e}nitra;$^{(c)}$Facult\'e des Sciences Semlalia, Universit\'e Cadi Ayyad, LPHEA-Marrakech;$^{(d)}$LPMR, Facult\'e des Sciences, Universit\'e Mohamed Premier, Oujda;$^{(e)}$Facult\'e des sciences, Universit\'e Mohammed V, Rabat;$^{(f)}$Institute of Applied Physics, Mohammed VI Polytechnic University, Ben Guerir; Morocco.\\
% ...
$^{46}$Physics Department, Southern Methodist University, Dallas TX; United States of America.\\
% ...
$^{51}$Fakult\"{a}t Physik , Technische Universit{\"a}t Dortmund, Dortmund; Germany.\\
% ...
$^{57}$II. Physikalisches Institut, Georg-August-Universit\"{a}t G\"ottingen, G\"ottingen; Germany.\\
% ...
$^{105}$CPPM, Aix-Marseille Universit\'e, CNRS/IN2P3, Marseille; France.\\
% ...
$^{120}$$^{(a)}$New York University Abu Dhabi, Abu Dhabi;$^{(b)}$United Arab Emirates University, Al Ain; United Arab Emirates.\\
% ...
$^{124}$Homer L. Dodge Department of Physics and Astronomy, University of Oklahoma, Norman OK; United States of America.\\
% ...
$^{156}$Raymond and Beverly Sackler School of Physics and Astronomy, Tel Aviv University, Tel Aviv; Israel.\\
% ...
$^{a}$ Also Affiliated with an institute covered by a cooperation agreement with CERN.\\
$^{b}$ Also at An-Najah National University, Nablus; Palestine.\\
$^{c}$ Also at Borough of Manhattan Community College, City University of New York, New York NY; United States of America.\\
$^{d}$ Also at Center for High Energy Physics, Peking University; China.\\
$^{e}$ Also at Center for Interdisciplinary Research and Innovation (CIRI-AUTH), Thessaloniki; Greece.\\
$^{f}$ Also at CERN, Geneva; Switzerland.\\
% ...

\end{flushleft}


%

%
%

%

%
%
%
%

\end{document}